# Modified Afshar Experiment: Calculations

Eduardo V. Flores
Department of Physics & Astronomy
Rowan University

**Abstract**

The Afshar experiment is a relatively simple two-slit experiment with results that appear to show a discrepancy with the predictions of Bohr's Principle of Complementarity. We report on the results of a calculation using a simpler but equivalent set-up called the modified Afshar experiment. Numerical results are in agreement with the experimental measurements performed on the Afshar experiment set-up. Calculations show that the level of which-way information and visibility in the Afshar experiment is higher than originally estimated.

## 1. Introduction

The Afshar experiment consists of coherent light incident onto a pair of pinholes [1,2]. The two emerging beams from the pinholes spatially overlap in the far-field and interfere to produce a pattern of alternating bright and dark fringes. At an appropriate distance from the pinholes thin wires are placed at the minima of the interference pattern. Beyond the wires there is a lens that focuses the light onto two photon detectors located at the image of each pinhole. When an interference pattern is not present, as in the case when only one pinhole is open, the wire grid obstructs the beam and produces scattering, thus reducing the total flux at the corresponding detector by about 14.38% [2]. However, when the interference pattern is present the disturbance to the incoming beams due to the wires is minimal, about 1%. From comparative measurements of the total flux with and without the wire grid, the presence of an interference pattern is inferred in a non-perturbative manner. Thus, the parameter $V$ that measures the visibility of the interference pattern is near its maximum value of 1.

When the wire grid is not present quantum optics predicts that a photon that hits a given detector originates from the corresponding pinhole with a very high probability. The parameter $K$ that measures the "which-way" information is 1 in this case. When a wire grid is placed at the dark fringes, where the wave-function is zero, the photon flux at the detectors hardly changes. We argue [1,2,3] that this is an indication that the wires have barely altered the "which-way" information, thus, $K$ is also nearly 1, which is in apparent violation of Englert's inequality, $V^2 + K^2 \leq 1$, a modern version of Bohr's principle of complementarity [4].

The modified Afshar experiment is a simpler and more transparent version of the Afshar experiment for calculation and analysis purposes [3]. A laser beam impinges on a 50:50 beam splitter and produces two spatially separated coherent beams of equal intensity (Fig.1). The beams overlap at some distance. Beyond the region of overlap the two beams fully separate again. There, two detectors are positioned such that detector 1 detects only the photons originating from mirror 1, and detector 2 detects only photons originating from mirror 2. Where the beams overlap they interfere forming a pattern of bright and dark fringes. At the center of the dark fringes we place thin wires.

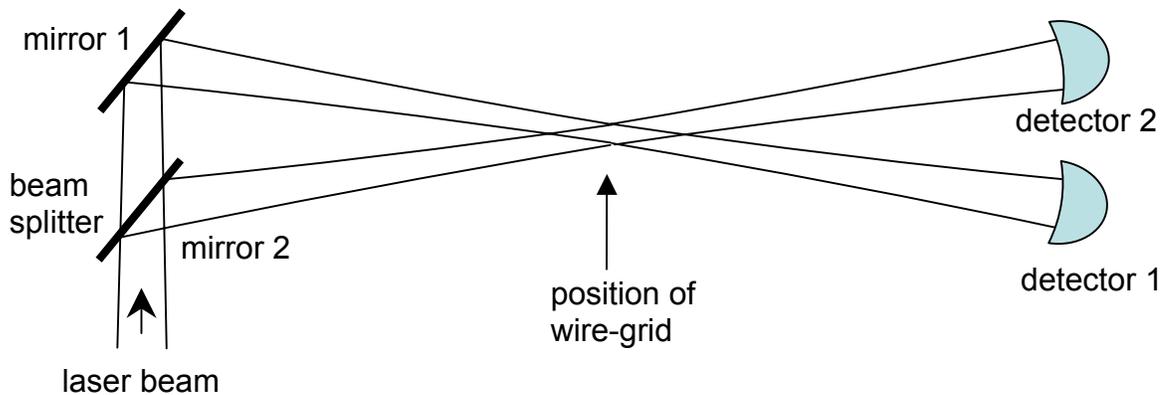

**Fig. 1** Modified Afshar experiment. The separation of the two beams occurs without an imaging system.

The modified Afshar experiment is equivalent to the Afshar experiment, except it does not use lenses. In particular, when the wire-grid is not present, the photons are free all the way from a mirror to its corresponding detector. Thus, the application of momentum conservation is straight forward. When a detector clicks, momentum conservation allows us to identify the particular path the photon took. Thus, we have full which-way information about the photon from the time it enters the interferometer until it hits the detector.

When the wire grid is in place it affects the path of the photon. To determine this effect we need to calculate wire diffraction. Once again, a wire diffraction calculation is simpler in the modified Afshar experiment where a uniform beam interacts with thin wires.

**2. Babinet's principle**

Diffraction of a laser beam by thin slits is one of the simplest calculations in optics. In our case we have thin wires instead of thin slits. Fortunately, the classical results for thin slits can be used in the modified Afshar experiment by a simple application of the principle of superposition also known as Babinet's principle [5]. We call $\vec{E}_0$ the electric field that arrives at the detection region when the laser beam is unperturbed, in other words, in the absence of the wire-grid. When the wire-grid is in place, the electric field that arrives at the detection region is $\vec{E}_W$. If instead of the wire-grid we use its complementary screen to disturb the laser beam then the electric field at the detection region is now $\vec{E}_S$. Babinet's principle states that $\vec{E}_0 = \vec{E}_S + \vec{E}_W$. In the case of the wire-grid its complementary screen is a thin-slits-grid where each slit is of the same thickness as a wire. In our calculations we use the Fraunhofer approximation since the distances from the sources and detectors to the diffracting object is a meter or more, which is relatively large compared to the size of the diffracting object, which is a millimeter or less. In the calculations below we assume a single polarization for light.

**3. The single beam case**

Since it is important for us to make contact with the experimental results obtained by Afshar et al., we first calculate wire diffraction for a single beam case. The single beam case is equivalent to having a single pinhole open. When pinhole A was open Afshar et al [2] observed a 14.14% reduction in the photon count at detector 1 and a deflection of 0.678% of the total number of photons to detector 2. When pinhole B was open and pinhole A was closed they reported a 14.62% decrease in photon count at detector 2. Unfortunately, they did not report the corresponding deflection to detector 1. We use their numbers to calibrate and test the modified Afshar experiment set-up. We use as free parameters the radius of the beam and the thickness of the wires. Afshar et al. use a ratio of about 10 to 1 for the center to center separation between wires to the thickness of a wire; we maintain the same ratio in the modified set up. In their experiment they use light with $\lambda = 638\,\text{nm}$, six wires with a thickness of $b = 128\,\mu\text{m}$, and their center to center wire separation is $d = 1.34\,\text{mm}$. In our calculation we use light with the same wavelength. We use six wires each with a thickness of $32\,\mu\text{m}$, the center to center wire separation is $319\,\mu\text{m}$ and the beam width is $3.22\,\text{mm}$.

The two laser beams that hit the slits propagate symmetrically on the *y-z*-plane and cross each other at the origin (Fig. 2). Each beam makes an angle $\alpha$ with the *z*-axis; $\alpha$ is small, 0.001 radians. The slit-grid is centered at the origin of the *x-y* plane. The long side of the slits is more than a centimeter long and it is parallel to the *x*-axis. Since the width of the beam is less than 5 mm no diffraction takes place along the *x*-direction. Diffraction takes place on the *y-z* plane. On this plane, diffraction is a function of the angle $\theta$ that diffracted light makes with the *z*-axis.

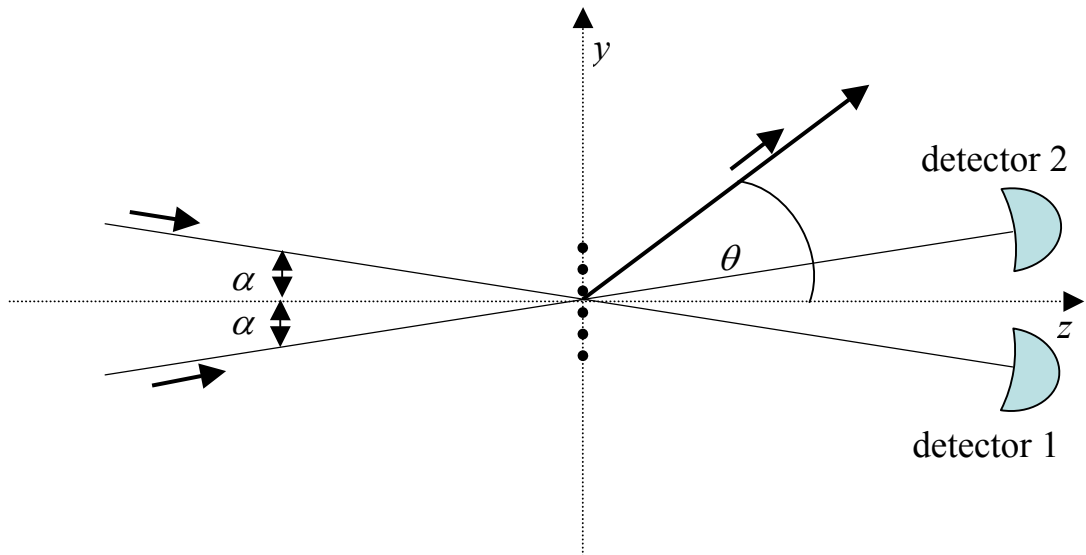

**Fig. 2** Two laser beams approach the wire-grid at a small angle $\alpha = 0.001\,\text{rad}$

with the *z*-axis. The wire-grid is centered at origin and lays on the *xy*-plane. Diffracted light makes an angle $\theta$ with respect to the *z*-axis. Two detectors are located right in front of each incoming beam.

The electric field produced by the interaction of a single beam with an opaque screen with *N* thin slits is a well known result given by: $E_S = \Lambda f(\theta)\sin(\omega t - \kappa R)$, where $\Lambda$ is a constant, $\omega$ is the angular frequency of light, $\kappa$ is the corresponding wave number, $t$ is time, $R$ is the distance from the center of the screen to the detector, and $f(\theta)$ is given by [6]

$$f(\theta) = \frac{1}{N}\left(\frac{\sin(\beta\sin(\theta))}{\beta\sin(\theta)}\right)\left(\frac{\sin(N\gamma\sin(\theta))}{\sin(\gamma\sin(\theta))}\right). \tag{1}$$

The thickness of the slit $(b)$, the center to center separation between slits $(d)$, and the magnitude of the wave vector $(\kappa)$ determine $\beta$ $(\beta = \kappa b/2)$ and $\gamma$ $(\gamma = \kappa d/2)$. To fully determine the electric field $E_S$ at the detection region we need to find $\Lambda$. We use energy conservation to determine this constant. In our case energy is proportional to the integral of the intensity over the area of the region under consideration. The energy that goes through the slits is equal to the energy that reaches the whole detection region. Thus, we find that $\Lambda$ is equal to $0.10751306 E_0$, where $E_0$ is the electric field of the unperturbed laser beam.

We use Babinet's principle to find the electric field, $E_W = E_0 - E_S$, produced by wire diffraction from the slit diffraction field $E_S$ and the unperturbed beam field $E_0$. In our case the electric field $E_0$ is zero everywhere except for a very small region where the detector in front of the unblocked mirror is placed. Where $E_0 = 0$ we have the useful relation $E_W(\theta) = -E_S(\theta)$, thus, the corresponding intensities produced by the wire-grid and the slit-grid are identical, $I_W(\theta) = I_S(\theta)$. We note that the calculation of the electric field $E_W$ at the detector in front of the original beam is complicated by the fact that the phase difference between of $E_S$ and $E_0$ is not a simple relation. Energy conservation is an alternative way to calculate the photon count at the detector in front of mirror.

Now that we have obtained the electric field $E_W$ we can calculate the intensity and predict the photon count relative to the unperturbed case. At the detection region two particular places are of great interest to us, the first being the detector in front of the original beam $(\theta = 0)$. We use energy conservation to calculate the percent photon count at this detector. We define $f$ as the photon count with respect to the photon count of the undisturbed beam times 100,

$$f = \frac{\text{photon count}}{\text{photon count undisturbed beam}} \times 100. \qquad (2)$$

The photon count for the undisturbed beam, $f_0$, is 100%. The calculation of the photon count losses at the wires, $f_W$, is simple; it involves an integral of the uniform intensity of light over the section of the wires that face the incoming beam; the result is 7.59199%. The photon count of light diffracted everywhere except at the detector in front of the incoming beam, $f_S$, is obtained from the integral of $I_S(\theta) \propto E_S^2$ over the whole detection region except at the detector in front of incoming beam; this value is 6.9652%. The equation of energy conservation is:

$$f_0 = f_W + f_S + f_D. \qquad (3)$$

The only unknown here is the photon count at the detector in front of incoming beam, $f_D$. Thus, we get $f_D = 85.4428\%$. This number corresponds to a 14.55% decrease in photon count which is in reasonable agreement with the experiment [2].

The second place of particular interest is the location of the other detector $(\theta = 2\alpha = 0.002)$ not in front of the incoming beam. The first diffraction peak happens to be right at the center of the second detector, $\theta = .002$ rad. Afshar et al reported that 0.678% of the photons in the original beam end up at their detector 2 [2]. We integrate $I_S(\theta)$ over the area of the second detector and predict that 0.627% of the photons of the original beam end up at the second detector. The relatively small discrepancy between the Afshar experiment and the modified Afshar experiment could be explained by the approximations made in the modified Afshar experiment to keep the calculations simpler such as the use of a beam with uniform profile and the use of the plane wave approximation. Also, there are experimental limitations such as the use of non-ideal lenses, imperfect alignments, and limitations of single photon detectors. On top of these issues, Afshar et al. did not have theoretical predictions at their disposal to guide their measurements.

**4. Two beam case**

The crucial calculation is wire diffraction when both beams are unblocked in the modified set-up. The calculation provides the information needed to determine the theoretical which-way information and visibility. The technique is similar to the one beam case. Thus, we first calculate the diffraction produced by a thin-slit-grid located at the minima of the interference pattern and then we use Babinet's principle to obtain the wire-grid diffraction.

We approximate the two beams as two plane waves with wave vectors $\vec{\kappa}_1 = -\hat{y}\kappa\sin\alpha + \hat{z}\kappa\cos\alpha$ and $\vec{\kappa}_2 = \hat{y}\kappa\sin\alpha + \hat{z}\kappa\cos\alpha$ (See Fig. 2). Superposition of the two plane waves $E_0 e^{i(\vec{\kappa}_1\cdot\vec{r}-\omega t)} + E_0 e^{i(\vec{\kappa}_2\cdot\vec{r}-\omega t)}$ results in a plane wave $E_{eff} e^{i(\vec{\kappa}'\cdot\vec{r}-\omega t)}$, where the wave vector $\vec{\kappa}' = \hat{z}\kappa\cos\alpha$ is directed perpendicular to the wire-grid, and $E_{eff}$ is an effective amplitude given by $E_{eff} = 2E_0 \cos(\kappa\sin(\alpha)y)$. Dark fringes appear around regions along the y-axis that fulfill condition: $\kappa\sin(\alpha)y = (2n+1)\pi/2$ for all integers $n$. We calculate the diffraction produced by a thin-slit-grid located at the center of the dark fringes of the interference pattern. Near the center of a dark fringe the effective amplitude is approximated by expanding $2E_0 \cos(\kappa\sin(\alpha)y)$ about this point. In this region the cosine function is linear with alternating slope from slit to slit. The result is $E_{eff} = \pm 2E_0 \kappa\sin(\alpha)u$, where $u$ is the distance from the center of the dark fringe.

For the particular case of a single thin slit at the center of a dark fringe the diffraction integral is [5]

$$\int_{-b/2}^{b/2} \frac{E_{eff}}{r} \sin(\omega t - \kappa r) du, \qquad (4)$$

where $b$ is the thickness of the slit, $r$ is the distance from a source point to the detector, and the effective amplitude is approximated by $E_{eff} = 2E_0 \kappa\sin(\alpha)u$. The distance $r$ is given by $r = (R^2 + u^2 - 2R\sin(\theta)u)^{1/2}$, where $R$ is the distance from the origin to the detector, and the angle $\theta$ is the angle that diffracted light makes with the z-axis. Since the phase is much more sensitive to small changes than the amplitude we may replace $r$ in the amplitude by $R$ but $r$ in the phase by $R - u\sin(\theta)$. Thus, the integral is now

$$\Omega \int_{-b/2}^{b/2} u \sin(\omega t - \kappa R + u\kappa \sin(\theta)) du, \qquad (5)$$

where $\Omega = \dfrac{2E_0 \kappa \sin \alpha}{R}$. Integrating this expression gives the electric field at the detector region as a function of $\theta$ for the single slit case. However, we need to calculate the effect of six slits.

The calculation of the electric field produced by 6 slits is a relatively simple extension of equation (5).

$$\sum_{j=0}^{j=5} (-1)^j \Omega \int_{jd-b/2}^{jd+b/2} (y - jd) \sin(\omega t - \kappa R + y\kappa \sin(\theta)) dy, \qquad (6)$$

where $d$ is the center to center distance between adjacent slits. The resulting electric field is a long expression that is easy to obtain with programs such as Mathematica. For our calculation we are more interested in the intensity. We calculate the intensity $I_S(\theta)$ by taking the time average of the square of the electric field,

$$I_S(\theta) = \frac{2\Omega^2}{(\kappa \sin \theta)^4} \{b\kappa \sin \theta \cos(1/2 b\kappa \sin \theta) - 2\sin(1/2 b\kappa \sin \theta)\}^2 \times$$

$$\{\sin(1/2 d\kappa \sin \theta) - \sin(3/2 d\kappa \sin \theta) + \sin(5/2 d\kappa \sin \theta)\}^2. \qquad (7)$$

A plot of the intensity $I_S(\theta)$ is presented in Fig. 3. We notice that this diffraction pattern predicted by the modified Afshar experiment calculation has been observed experimentally using the Afshar experiment set-up [6]. This is an additional confirmation of the equivalence of the two set-ups.

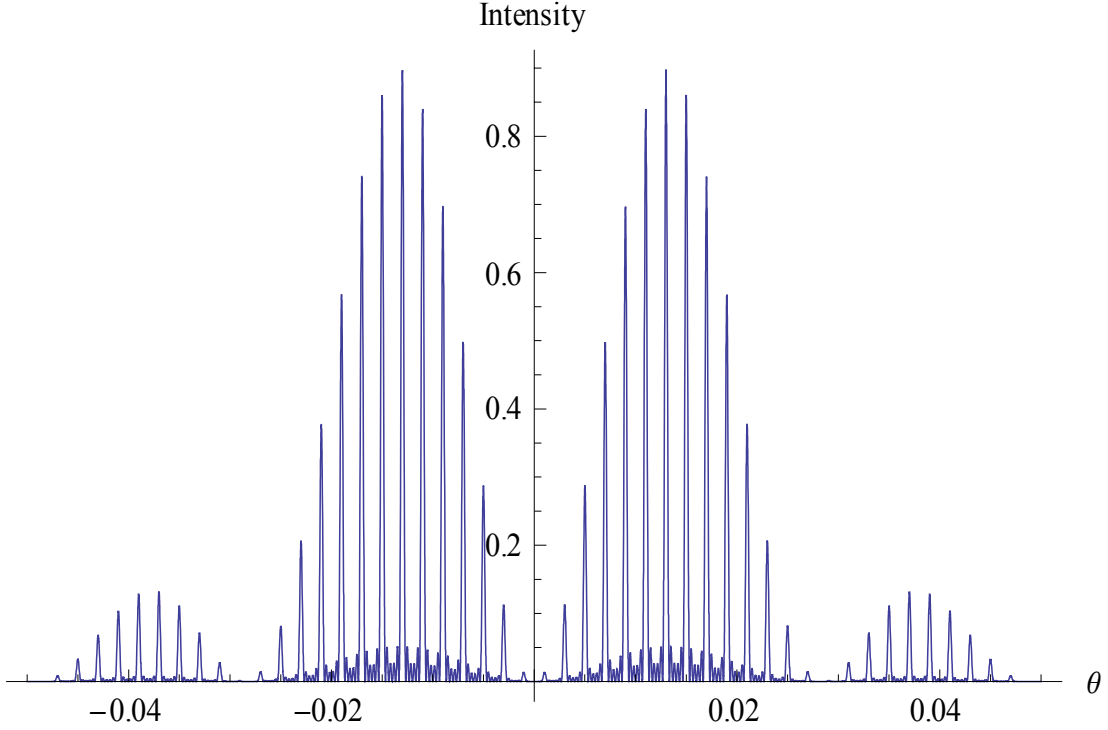

**Fig. 3** A plot of the intensity $I_S(\theta)$ for the case when two coherent beams interfere and then diffract due to a thin-slit-grid located at center of dark fringes. Notice that detectors 1 and 2 are symmetrically located right at the first peak, $\theta = \pm 0.001\,\text{rad}$. The intensity at the first peak is hardly noticeable in the graph.

We calculate the percent photon count at either detector, $f_D$. For this purpose we use a conservation of energy equation similar to equation (3):

$$f_0 = f_{WG} + f_{LD} + f_D \qquad (8)$$

The photon count for the undisturbed beam, $f_0$, is 100%. The decrease in photon count at the wire-grid, $f_{WG}$, is proportional to the integral of the square of the effective amplitude, $E_{eff} = 2E_0 \kappa \sin(\alpha) u$, at the wires. We get $f_{WG} = 0.125667\,\%$. The photon count of light diffracted everywhere except at the detectors, $f_{LD}$, is obtained from the integral of $I_S(\theta)$ in equation (7) over the whole detection region except at the detector; we find that $f_{LD} = 0.124709\,\%$. The only unknown in equation (8) is the photon count at the detector in front of incoming beam, $f_D$. Thus, we get $f_D = 99.7496\,\%$. Therefore, the percent decrease in photon count at either detector is 0.25%. Afshar et al. reported that the percent decrease in photon count at their detector 1 was 0.31% [2]. Our calculation is in reasonable agreement with this measurement. They also reported that their detector 2 showed a decrease in photon count of 1.13% which is high compared with our calculation. The discrepancy is understandable considering that they did not have a theoretical calculation to guide their findings and that their experiment is quite sensitive to correct alignment.

**5. The which-way information**

Diffracted light has no which-way information since it could come from either mirror. To estimate the which-way information we need to know the amount of diffracted light that reaches the detectors. The intensity $I_S(\theta)$ in equation (7) shows that most of the diffracted light falls away from the detectors. In fact in Fig. 3 we can see that the location of the detectors is at the very first peak of both sides of the pattern where the intensity is hardly noticeable. Integrating $I_S(\theta)$ in equation (7) over the detector area allows us to get the fractional photon count of diffracted light that reaches the detector, $9.58447 \times 10^{-6} \approx 1/100,000$. Now, the situation seems clearer to us. Consider 100,000 particles that come from a given mirror towards its corresponding detector one at a time. The wires stop 126 of these particles. The total number of diffracted particles is also 126. Of these diffracted particles, 125 fall outside the detector. Only 1 particle is diffracted to a detector. Since this particle has been diffracted it has no which-way information. The remaining 99,748 particles have which-way information; they come directly from the mirror to its corresponding detector. Thus, the which-way information parameter $K$ is close to 1.

**6. Visibility**

Unfortunately, we cannot measure the visibility directly but we can place a lowest limit [2,7]. To calculate the lowest limit for the visibility we use the fact that out of 100,000 photons that go from a mirror towards a given detector 126 are stopped by the wires and the remaining go through. We use all the photons available to provide the lowest limit for the visibility. The photons that are stopped by the wires must be part of the minimum intensity region while the photons that go through must be part of the high intensity region.

We start by assuming ignorance about the shape of the interference pattern. We consider the standard formula for the visibility

$$V = \frac{I_{max} - I_{min}}{I_{max} + I_{min}} \qquad (9)$$

where $I_{max}$ and $I_{min}$ are the maximum and minimum intensities of the interference pattern. To minimize the visibility $I_{max}$ needs to be as small as possible and $I_{min}$ as large as possible. To maximize $I_{min}$ the darker regions must have the geometrical shape of thin rectangular boxes each with a base equal to the thickness of the wire, 0.032 mm, times its corresponding length $\approx 3.22$ mm. $I_{min}$ is proportional to 126 divided by the exact area of the base of the six thin boxes, 0.5801 mm$^2$, $I_{min} \propto 217.2$. Similarly, $I_{max}$ is minimized by distributing uniformly the photons that miss the wires (99,874) on an area equal to the beam cross section subtracting the area covered by the wires, 7.56322 mm$^2$; the result is $I_{max} \propto 13,205.2$. Thus, the interference pattern with the lowest visibility compatible with our data is a type of periodic square function (Fig. 4). Using equation (9) we get the lowest limit for the visibility, $V \geq 0.968$. Our lowest limit is higher than the one obtained by Afshar et al. [2,7]. This is so because our calculation gives us the number of photons stopped by the wire, Afshar et al. did not measure this number.

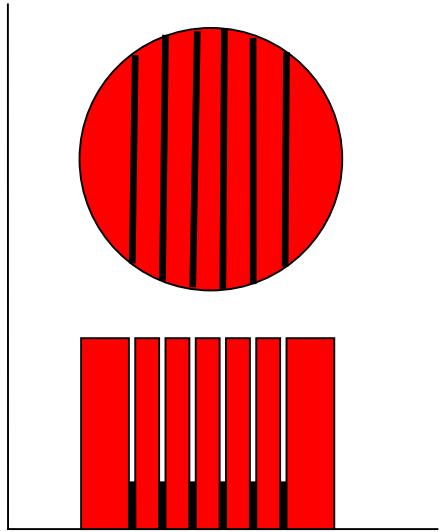

**Fig. 4** Two views of interference pattern with lowest visibility at the location of the wire-grid. Low intensity bars $I_{min}$ are maximized by uniformly distributing the photons that stay at the wires, 0.126%, over the cross sectional area of the wires. High intensity bars $I_{max}$ are minimized by uniformly distributing the photons that go through, 99.874%, over the remaining beam cross section. The visibility of this interference pattern is $V = 0.968$.

## 7. Conclusion

In the modified Afshar experiment we have set a lowest limit for the visibility of the interference pattern, $V \geq 0.968$. Thus, we have evidence of a sharp interference pattern at the location of the wires. Interference is a reliable indicator of the wave aspect of the photon. Similarly, a calculation shows that the which-way information parameter $K$ is very high, $\approx 1$. Which-way information about the path of the photon can be associated to the particle aspect of the photon. Squaring the parameters $K$ and $V$ and adding them we get $V^2 + K^2 \geq 1.93$; a clear evidence of the coexistence of particle and wave beyond the limitations imposed by complementarity.

Englert analyzed two-way-interferometer experiment such as a Mach-Zehnder setup [4]. Englert's main results in his derivation of the duality relation, $V^2 + K^2 \leq 1$, is that this relation for is independent of the uncertainty principle. It is easy to see why the modified Afshar experiment is not bound by the uncertainty principle. The which-way information and visibility are obtained from the decrease in photon count at the detectors and from the photons stopped by the wires respectively. These two numbers are related but they are no two conjugate variables to form an uncertainty relation of the Heisenberg-Roberson kind. Thus, it appears that the violation of the duality relation in the modified Afshar experiment is in no way a violation of the uncertainty principle.

We notice that Englert's duality relation, $V^2 + K^2 \leq 1$, was obtained from wave or matrix mechanics. Thus, wave or matrix mechanics alone cannot provide a way to violate this inequality relation. This is understandable since quantum mechanics shows the development of the wave function not of point particles. Therefore, in order to break complementarity we need an external factor to bring out the particle aspect of the photon. It turns out that for the modified Afshar experiment the click of a detector together with the photon momentum is the right kind of external factor. Due to momentum conservation, the photon momentum is a faithful marker even when the visibility is near 1. The calculation shows that only 1 in 100,000 photons that hit a detector may come from the wrong mirror in the modified Afshar experiment. Thus, momentum conservation allows us to claim with high confidence the likely path of the photon backwards from the detector to its corresponding mirror. We note that this technique, detector click plus momentum conservation, has been used before to find which-way information in the delay choice experiment [8].

**Acknowledgements**

The author would like to thank Prof. Ernst Knoesel, Scott Roszko, and Gabriela Hristescu for their help with this work.